\documentclass[aps,prd,nofootinbib,onecolumn]{revtex4}
\usepackage{graphicx,pstricks,rotating}
\usepackage{amsmath}
\usepackage{graphics}
\usepackage{epsfig}
\begin{document}
\title{
Two  photon decays of scalar mesons in the quark NJL model
}

\author{Yu. L. Kalinovsky}
\email{kalinov@jinr.ru}
\affiliation{%
Laboratory of Information Technologies, Joint Institute for Nuclear Research,
141980 Dubna, Russia }

\author{M. K. Volkov}
\email{volkov@theor.jinr.ru}
\affiliation{%
Bogoliubov Laboratory of Theoretical Physics, Joint Institute for Nuclear Research,
141980 Dubna, Russia }

\begin{abstract}
Two  photon decays of
scalar mesons $f_0(980)$, $a_0(980)$, $\sigma(600)$
in the quark Nambu - Jona - Lasinio (NJL) model
are calculated.
The contributions of the  meson loops
are  taken into account  along with the quark loops (Hartree - Fock approximation).  
This corresponds to the next order of  the $1/N_c$ expansion, where
 $N_c=3$ is the number of quark colors. It is shown that the meson  and quark loops give
  comparable  contributions to the amplitude. Moreover, in the process  $f_0(980) \rightarrow \gamma \gamma$
 the kaon loop plays the dominant role. A similar situation takes place in the decay
 $\phi \rightarrow f_0(980) \gamma$\cite{physrev}.
 Our results are in satisfactory agreement with the recent experimental data.
 \end{abstract}

\maketitle


\section{Introduction}

Experimental studies of  two  photon decays of  scalar mesons
play an important role in testing different theoretical models.
In the last few years a number of  experimental
investigations have been devoted to this problem \cite{a1,a2,a3,a4}.

There is a number of models which attempt to explain  the
inner structure of these scalar mesons  and their interactions.
One of them considers the scalar mesons as  four - quark states\cite{achasov,achasov2}.
Another model interprets these mesons as an  admixture of 
quark - antiquark  and diquark - antidiquark  states\cite{gerasimov}.
There is a model which describes the scalar states as  kaon molecules\cite{kaon1}-\cite{kaon4}.

Here we use the  quark NJL model\cite{NJL,echaya}. This model allows us to successfully
describe the low - energy hadron physics using the chiral symmetry of the strong  interaction
\cite{echaya,NJL2,echaya2}.
As a rule, in this model the Hartree - Fock approximation is used
for consideration of the quark loops only.
However, there are some  processes, for instance, the radiative decays with participation of
scalar mesons, where the meson loops can also play a very important role\cite{physrev,feldmann}.
Consideration of the meson loops implies allowance for
the next order to the $1/N_c$ approximation in the NJL model.
Recently,  it has been  shown  that in the process $\phi \rightarrow f_0(980) \gamma$  the kaon  loop
plays even the dominant role as compared to the quark loop\cite{physrev}.  In this case the kaon loop 
 completely defines the decay width of this process in  agreement with the experimental data.
 
Our paper is devoted to study of the two  photon decays of the scalar mesons $f_0(980)$, $a_0(980)$ and $\sigma(600)$.
It is a natural continuation of the investigations fuilfiled in \cite{physrev}.
Here we also show  that in the decay $f_0(980) \rightarrow \gamma \gamma$ the kaon loop
again plays the dominant role.

The obtained results  are in satisfactory agreement with the recent experimental data.

In the next section,  we present  part of the meson - quark Langrangian
obtained  from the  quark  NJL model \cite{echaya}. The electromagnetic  interaction
is introduced by the standard method.
We describe two  photon decays of
the scalar mesons $f_0(980)$, $a_0(980)$, $\sigma(600)$
with the help of this Lagrangian.

In section III, we calculate two  photon decays of  scalar mesons taking into account
the contributions of the quark and meson loops.

In conclusion, we shortly  discuss the obtained results.

 \section{The NJL model.}
The part of the Lagrangian which we need for the description
 of the two photon decays of the scalar mesons
 has the form\cite{echaya}

 \begin{eqnarray}\label{lagrangian}
{\cal L}&=& \bar{q} \Bigl\{  i \hat{\partial}  -  {m} + e Q  \hat{ A}   \nonumber \\
&+&  g_{\sigma_u} \lambda_u  \sigma_u  +  g_{\sigma_s}\lambda_s \sigma_s
+  g_{\sigma_u} \lambda_3 a_0 + \nonumber \\
&+& i \gamma_5 \Bigl[  g_\pi  \left(  \lambda_{\pi^+}\pi^+  + \lambda_{\pi^-}\pi^-  \right)
\nonumber \\
&+ & g_K  \left( \lambda_{K^+}K^+ +\lambda_{K^-}K^- \right) \Bigr]
 \Bigr\} q.
\end{eqnarray}

Here ${q}=({u},{d},{s})$ are the  quark fields with the masses 
$m=\mbox{diag}(m_u, m_d, m_s)$;  $m_u,m_d,m_s$ are the constituent quark masses ($m_u=m_d$); 
$A_\mu$ are the photon fields, $e$ is the electric charge,
$Q=(\lambda_3+\lambda_8/\sqrt{3})/2$. 
$\pi^-,\pi^+, K^-, K^+$ are the pseudoscalar  mesons; 
$ a_0(980)$ is the isovector scalar meson with the mass  $980$ MeV;  
$\sigma_u, \sigma_s$ are components of isoscalar scalar mesons $\sigma (600), f_0(980)$. 
The state $\sigma_u$ consists of light $u$ and $d$ quarks, and the state $\sigma_s$ 
consists of $s$ quarks  only. 
The physical scalar mesons
$\sigma (600), f_0(980)$
are expressed in terms of $\sigma_u, \sigma_s$  with the help of
the angle $\alpha =\theta_0-\theta$. Here $\theta_0=35.3^\circ$  and $\theta =24^\circ$ are the ideal
and the real singlet - octet mixing angles\cite{echaya,angle}
\begin{eqnarray}
&& \sigma = \sigma_u \cos \alpha -\sigma_s \sin \alpha, \nonumber \\
&& f_0 = \sigma_u \sin \alpha +\sigma_s \cos \alpha.
\end{eqnarray}

The matrices $\lambda_i$ are expressed in terms of the Gell - Mann matrices:
$\lambda_u = (\lambda_8+\sqrt{2} \lambda_0)/\sqrt{3}$,
$\lambda_s = (-\lambda_0+\sqrt{2} \lambda_8)/\sqrt{3}$,
$\lambda_{\pi^{\pm}}=(\lambda_1\mp i\lambda_2)/\sqrt{2}$,
$\lambda_{K^{\pm}}=(\lambda_4\mp i\lambda_5)/\sqrt{2}$, $\lambda_0=\sqrt{2/3}\,\,{\tt 1}$.
The coupling constants can be written in terms of  the logarithmic divergent integral as
(see \cite{echaya})
\begin{eqnarray}\label{cons}
&& g_{\sigma_u}=\left( 4 I^\Lambda(m_u, m_u)\right)^{-1/2}, \,\,
 g_{\sigma_s}=\left( 4 I^\Lambda(m_s, m_s)\right)^{-1/2},  \nonumber \\
&&  g_{\pi} = Z_{\pi} g_{\sigma_u}, \,\,
g_K = Z_K \left( 4 I^\Lambda(m_u, m_s)\right)^{-1/2},
\end{eqnarray}
where
\begin{eqnarray}\label {integral}
I^\Lambda(m_1, m_2) &=& \frac{N_c}{(2\pi)^4} \int d^4 k \frac{\theta(\Lambda^2-k^2)}{(k^2+m_1^2)(k^2+m_2^2)}
\nonumber \\
&=& \frac{3}{(4\pi)^2 }
\Bigl[
m_2^2 \ln \left( \frac{\Lambda^2}{m_2^2}+1\right)
\nonumber \\
& & -
m_1^2 \ln \left( \frac{\Lambda^2}{m_1^2}+1\right)
\Bigr]/(m_2^2-m_1^2).
\end{eqnarray}
All these integrals are presented  in the Euclidean space.
The constituent quark masses are $m_u = m_d = 260$ MeV,
$m_s = 410$  MeV   and the cutoff parameter  $\Lambda = 1.27$ GeV
\cite{physrev,echaya,volkovlast}
\footnote
{Let us note that in \cite{echaya} some other values of these parameters were used
which corresponded to the weak pion decay constant $F_\pi = 93$ MeV and   the width
$\Gamma_{\rho \rightarrow \pi\pi} = 155$ MeV. Here we use modern experimental data
\cite{a1} for fixing our model parameters:  $F_\pi = 92.4$ MeV and  
$\Gamma_{\rho \rightarrow \pi\pi} = 149.4$ MeV.}.
As a result,  we obtain  $g_{\sigma_u} = 2.42$, $g_{\sigma_s} = 3.0$; 
$Z_\pi$ and $Z_K$ are the factors which take into account the transitions of 
pseudoscalar mesons to axial - vector mesons.   
$Z_{\pi} = (1 -  \frac{6 m_u^2}{M_{a_1}^2} )^{-1/2}$,
$Z_K=(1-\frac{3(m_u+m_3)^2}{2M_{K_1}^2})^{-1/2}$
 \cite{echaya}, where
$M_{a_1}=1260$ MeV,  $M_{K_1}=1403$ MeV are the masses of the axial -vector mesons\cite{a1}.
In our calculations we  suppose $Z_\pi \approx Z_K = Z = 1.2$.
As a result,  we have
$g_\pi \approx 2.9$, $g_K \approx 3.3$. 

\section{Two  photon decays of  scalar mesons.}

\begin{figure}
\includegraphics[scale=1]{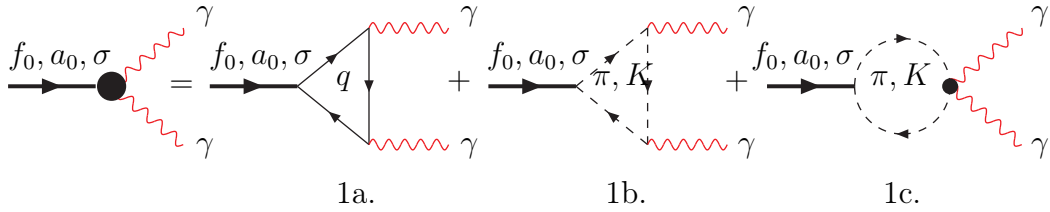}
\caption{\label{figure1}
Loop diagrams for the two photon decays of the scalar mesons.
The diagram 1a. corresponds to processes going through the quark loops
(Hartree - Fock approximation), $q=u,d,s$. The diagrams 1b. and 1c.
define processes connected with the meson loops (next order of the $1/N_c$ expansion). 
}
\end{figure}

Two photon  decays of  scalar mesons are described by the diagrams in Fig. \ref{figure1},
where the first diagram 1a.
determines the contribution of  the quark loops (Hartree - Fock approximation),
and the other two (1b.,1c.) describe  the contribution of  the meson loops (next order of the $1/N_c$ expansion). 
The strong vertices in the first diagram are given in the Lagrangian (\ref{lagrangian}).
The strong vertices in the meson loops  are defined by the divergent parts of the quark loops
(see eqs. (\ref{cons}) and (\ref{integral})). 
It is a standard method of the local NJL model\cite{physrev,echaya,feldmann}.

 The amplitude of the two  photon decays of the scalar (S) mesons  has the  form:
 \begin{eqnarray}
T_S^{\mu\nu} = \frac{\alpha}{\pi F_\pi}
(A_S^{q}+A_S^{M}) \left( g^{\mu\nu} (q_1q_2) -q_1^{\nu} q_2^{\mu} \right), 
\end{eqnarray}
 where $A^q_S$ is the quark loop contribution and $A^M_S$ is the meson loop contribution, 
 $q_1, q_2$ are the photon momenta, $\alpha = e^2/4\pi=1/137$.

The contributions to the amplitude $T_S^{\mu\nu}$ from the quark loops are 
 calculated in detail in \cite{echaya,echaya2}. In the $q^2$  -  expansion  over photon momenta
they have the following form:

\begin{eqnarray}
&& A_{a_0(980)}^u = \frac{2}{3Z} = 0.48,  \\
&&  
A_{\sigma(600)}^{u} = \frac{10}{9Z}  \cos \alpha = 0.8 , \,\, 
A_{\sigma(600)}^{s} = \frac{2\sqrt{2}}{9Z} \frac{F_\pi}{F_s} \sin \alpha  = ,
0.24 \\
&& 
A_{f_0(980)}^{u}=  \frac{10}{9Z}  \sin \alpha  = 0.6, \,\, 
A_{f_0(980)}^{s}= - \frac{2\sqrt{2}}{9Z} \frac{F_\pi}{F_s} \cos \alpha  = - 0.35 .
\end{eqnarray}
 Here  we have used the Goldberger - Treiman relation for the constants
$g_{\sigma_u}$ and $g_{\sigma_s}$:
$g_{\sigma_u} = g_\pi /Z = m_u /(ZF_\pi)$,
$g_{\sigma_s} =  m_s /(ZF_s)$,  where $F_s=1.28 F_\pi$ \cite{echaya,echaya2}.


This approximation allows us  to obtain, in the framework of the local NJL model, 
the Wess - Zumino terms in the phenomenological chiral Lagrangian
\cite{deffen}.  With help of these  terms
it is possible to succesfully describe main radiative meson decays with participation of pseudoscalar and vector mesons
such as 
$(\pi^0, \eta, \eta') \rightarrow \gamma\gamma $,
$(\rho, \omega) \rightarrow (\pi, \eta) \gamma$,
$K^* \rightarrow  K \gamma$, $\eta' \rightarrow (\rho, \omega) \gamma$,
$\phi \rightarrow (\eta, \eta') \gamma$ \cite{echaya,echaya2} 
\footnote{Let us note that  in \cite{feldmann}  some other method for calculation of the
quark loops was used. Here we do not use this method, because  it does not allow us
to obtain the agreement with the experimental data for the above-mentioned  radiative  decays 
with pseudoscalar and vector mesons.
The method in \cite{feldmann} also has  the problem with the quark confinement.
Let us note that we use a different value for the singlet - octet mixing angle $\theta$ obtained with  the 
help of taking into account  the  t' Hooft interaction\cite{angle}.
}.



The strong meson vertices describing the two photon decays of the scalar mesons  
(see Fig. \ref{figure1})  have the form




\begin{eqnarray}
&& G_{a_0(980)\pi^+\pi^-} = 0 , \,\,  
G_{a_0(980)K^+K^-} = 2  (2m_u-m_s)\frac{g_K^2}{g_{\sigma_u}}, \\
&&
G_{f_0(980) \pi^+ \pi^-} = 4  m_u \frac{g_\pi^2}{g_{\sigma_u}} \sin \alpha \\
&& G_{f_0(980) K^+K^- } =
2\left[ \sqrt{2} (2m_s-m_u) \frac{g_K^2}{g_{\sigma_s}}   \cos \alpha
- (2m_u-m_s) \frac{g_K^2}{ g_{\sigma_u} }   \sin \alpha  \right] \\
&&G_{\sigma(600) \pi^+ \pi^-} = 4 {m_u}  \frac{g_\pi^2}{g_{\sigma_u}}\cos \alpha , \\
&&G_{\sigma(600) K^+ K^-} =
2\left[ \sqrt{2} (2m_s-m_u) \frac{g_K^2}{g_{\sigma_s}}   \sin \alpha
+ (2m_u-m_s) \frac{g_K^2}{g_{\sigma_u}}   \cos \alpha  \right] 
\end{eqnarray}

The sum of diagrams 1b. and 1c.  leads to the converged integral having  the 
gauge invariant form. 

These integrals were caclulated in the 
analytical form in\cite{feldmann}.  
Particularly, in the decay $a_0(980) \rightarrow \gamma \gamma$ the contribution $A^{K^+}_{a_0}$
of the kaon loop
to the amplitude of this process
 is expressed in terms of the known function  $\varphi(x)$

\begin{eqnarray}
A^{K^+}_{a_0} = (2m_u-m_s) \frac{g_K^2}{g_{\sigma_u}}
\frac{F_\pi}{M_{a_0}^2} (x_1 \varphi(x_1) -1)
\end{eqnarray}
where
\begin{eqnarray}
\varphi (x) =
\left\{
\begin{array}{l}
\left[ \arctan(x-1)^{1/2} \right]^2,  x \geq 1 \\
 \left[ \frac{i}{2} {\ln}
 \frac{1-\sqrt{1-x}}{1+\sqrt{1-x}} - \frac{\pi}{2}
 \right]^2, x  \leq 1
\end{array}
\right. ,
\end{eqnarray}
and $x_1=4M_K^2/M_{a_0}^2$.
This leads to the numerical value
$A^{K^+}_{a_0} \approx -0.1$.

For the total amplitude we have 

\begin{eqnarray}
T_{a_0}^{\mu\nu} = \frac{\alpha}{\pi F_\pi} (0.48-0.1)( g^{\mu\nu}(q_1q_2)-q_1^\nu q_2^\mu ) 
= \frac{\alpha}{\pi F_\pi} 0.38 (g^{\mu\nu}(q_1q_2)-q_1^\nu q_2^\mu) 
\end{eqnarray}

For the decay width we obtain
\begin{eqnarray}
\Gamma_{a_0\rightarrow \gamma \gamma } =
\frac{M_{a_0}^3}{64\pi}|T_{a_0\rightarrow \gamma \gamma}|^2 = 0.39  \, \mbox{keV}.
\end{eqnarray}
The experimental value is $ \Gamma(a_0(980) \rightarrow \gamma \gamma) =0.3^{+0.11}_{-0.10} $ keV\cite{a1}.

For the decays $f_0(980) \rightarrow \gamma \gamma$ and
$\sigma(600) \rightarrow \gamma \gamma$ the meson contributions
to the amplitudes have the form
\begin{eqnarray}
&&A^{\pi}_{f_0(980)}   
= 
4  \frac{m_uf_\pi}{M_{f_0}^2}
 \frac{g_\pi^2}{g_{\sigma_u}} ({x_2} \varphi(x_2) -1)_{\pi}\sin \alpha 
=  0.07 + i 0. 18
 \\ 
&& A^K_{f_0(980)}  
= 
2\sqrt{2}  \frac{(2m_s-m_u) F_\pi}{M_{f_0}^2}
 \frac{g_K^2}{g_{\sigma_s}} ({x_3} \varphi(x_3) -1)_{K}\cos \alpha  \nonumber \\
&& - 2  \frac{(2m_u-m_s) F_\pi}{M_{f_0}^2}
 \frac{g_K^2}{g_{\sigma_s}} ({x_3} \varphi(x_3) -1)_{K}\sin \alpha  = 0.536 \\ 
\end{eqnarray}
where $x_2=4M_\pi^2/M_{f_0}^2$, $x_3=4M_K^2/M_{f_0}^2$. 
The total amplitude  is
\begin{eqnarray}
T_{f_0}^{\mu\nu} = \frac{\alpha}{\pi F_\pi} (0.5-0.35 - (  0.07 + i 0. 18) 
- 0.536)( g^{\mu\nu}(q_1q_2)-q_1^\nu q_2^\mu ) 
= \frac{\alpha}{\pi F_\pi} (-0.23- i 0.18) (g^{\mu\nu}(q_1q_2)-q_1^\nu q_2^\mu) 
\end{eqnarray}
Then for the decay width we obtain 
$\Gamma (f_0 \rightarrow \gamma \gamma) = 0.19$ keV.  
There are several experimental data for the decay  $\Gamma(f_0(980)\rightarrow \gamma \gamma )$:
$\Gamma(f_0(980)\rightarrow \gamma \gamma ) = 0.29^{+0.07}_{-0.09}$ keV \cite{a1},
$\Gamma(f_0(980)\rightarrow \gamma \gamma ) = 0.205^{+0.095\,+0.147}_{-0.083\,-0.117}$ keV \cite{a2},
 $\Gamma(f_0(980) \rightarrow\gamma \gamma ) = 0.31\pm0.14\pm0.09$ keV \cite{a3},
 $\Gamma(f_0(980) \rightarrow\gamma \gamma ) =  0.29\pm0.07\pm0.12$ keV \cite{a4}.
  In other models the following results were obtained:
  $\Gamma(f_0(980)\rightarrow \gamma \gamma ) = 0.29^{+0.07}_{-0.09}$ keV \cite{a1},
  $\Gamma(f_0(980) \rightarrow\gamma \gamma ) = 0.24$ keV\cite{ivanov},
   $\Gamma(f_0(980)\rightarrow \gamma \gamma ) = 0.28^{+0.09}_{-0.13}$  keV\cite{anisovich},
  $\Gamma(f_0(980) \rightarrow\gamma \gamma ) = 0.31$  keV \cite{scadron},
  $\Gamma(f_0(980)\rightarrow \gamma \gamma ) = 0.33$  keV\cite{schumacher},
  $\Gamma(f_0(980) \rightarrow\gamma \gamma ) = 0.27$  keV\cite{achasov1},
  $\Gamma(f_0(980) \rightarrow\gamma \gamma ) = 0.20$  keV\cite{oller},
  $\Gamma(f_0(980)\rightarrow \gamma \gamma ) = 0.22 \pm 0.07$  keV\cite{hanhart},
  $ \Gamma(f_0(980) \rightarrow\gamma \gamma ) = 0.21- 0.26$ keV\cite{kaon4}.

For the decay amplituide  $\sigma(600) \rightarrow \gamma \gamma$ we have
\begin{eqnarray}
 A^{\pi}_{\sigma(600)}  
=  
4  \frac{m_uf_\pi}{M_{\sigma}^2}
 \frac{g_\pi^2}{g_{\sigma_u}} ({x} \varphi(x) -1)_{\pi}\cos \alpha  = 
-0.83 + i 0.83
 \end{eqnarray}
 where $x=4M_\pi^2/M_\sigma$. 
 \begin{eqnarray}
&& A^K_{\sigma(600)} 
= 
2\sqrt{2}  \frac{(2m_s-m_u) F_\pi}{M_{\sigma}^2}
 \frac{g_K^2}{g_{\sigma_s}} ({x_4} \varphi(x_4) -1)_{K}\sin \alpha  \nonumber \\
&& + 2  \frac{(2m_u-m_s) F_\pi}{M_{\sigma}^2}
 \frac{g_K^2}{g_{\sigma_s}} ({x_4} \varphi(x_4) -1)_{K}\cos \alpha 
\end{eqnarray}

Keeping  together all contributions to the amplitudes  we have the following results:
$A^{u,s,\pi^+,K^+}_{f_0} = 0.22$, 
$A^{u,s,\pi^+,K^+}_{\sigma}  = 1.6+i 0.8$ 
and the corresponding widths are
$\Gamma(\sigma \rightarrow \gamma \gamma) = 1.03$ keV.

  Unfortunately, for the decay $\Gamma(\sigma \rightarrow \gamma \gamma)$ 
  there are no reliable 
  experimental data. In other theoretical models the following estimations were
  obtained:  
  $\Gamma (\sigma \rightarrow \gamma \gamma) = 0.45$ keV \cite{achasov2}.
  $\Gamma (\sigma \rightarrow \gamma \gamma) < 1$ keV if $M_\sigma < 0.7-0.8$ MeV\cite{valera2}.
  $\Gamma (\sigma \rightarrow \gamma \gamma) = 4.1 \pm 0.3$ keV, $M_\sigma = 441$ MeV\cite{oller2}.

We see that our results do not  contradict  the recent experimental data. 
Let us note that if we  use a smaller  mass of $\sigma$  meson,  we will obtain the result 
close to the prediction \cite{achasov2,valera2}.

\section{Conclusion}

As a result, we have shown that in the two - photon decays of $f_0(980), a_0(980), \sigma(600)$
the quark and meson loops give  comparable  contributions to the amplitude of these processes.

In the decay $f_0(980) \rightarrow \gamma \gamma $ the kaon loops play even the dominant role.
Indeed, in this process
the  contributions from the $u(d)$ and $s$ quark loops noticeably cancel each other. Therefore,
the contributions of meson loops became dominant. Some other reasons lead to a similar
result in the process   $\phi \rightarrow f_0(980) \gamma$ \cite{physrev}.

On the other hand, in the processes $(a_0(980),\sigma(600)) \rightarrow \gamma \gamma $
the contributions from the quark and meson loops have the same order. However, the
main contribution to these amplitudes is connected with the quark loop.

Our results concerning the processes  $f_0(980) \rightarrow \gamma \gamma $ and
$\phi \rightarrow  f_0(980) \gamma$ do not
contradict  the theoretical predictions obtained  in other models where the full
contributions were defined  by the  kaon loops.

Let us note that the amplitudes
describing the radiative decays of scalar mesons   contain two different parts:
the first is connected with the quark intermediate state (quark loop);  and the second, with 
 the four-quark intermediate state (meson loop).   Therefore,  these amplitudes
 can be  considered   as a mixing of the two quark and four quark intermediate hidden states.

 On the other hand, the radiative decays with pseudoscalar mesons (see section 3.) are totally defined
 by the  quark loops only.  This fact confirms the ($\bar{q}q$)  structure of the pseudoscalar
 mesons.

In conclusion, we would like to emphasize that in the framework of our model we did not use any additional parameters 
for the description of radiative decays with participation of scalar mesons.

{\bf Acknowledgements} We thank  Profs.
N. N. Achasov, D. Blaschke, S. B. Gerasimov, D. Ebert and  V. N. Pervushin for  fruitful discussions.
This work was supported in part (Yu. K.) by the RFFI Grant No. 06-01-00228.

\end{document}